\documentclass[9pt,technote]{IEEEtran}

\usepackage[numbers,sort&compress]{natbib}
\usepackage{natbib}
\usepackage{url}
\usepackage[]{hyperref}
\usepackage{cleveref}
\usepackage{verbatim}
\usepackage{amsmath,amssymb}
\usepackage{rotating}
\usepackage{tikz}

\author{Aaron F. McDaid, Derek Greene, Neil Hurley \\ Clique Reseach Cluster, University College Dublin, Ireland. \\ {\tiny \textit{aaronmcdaid@gmail.com}}}
\title{Normalized Mutual Information to evaluate overlapping community finding algorithms}
\begin{document}

\maketitle

\newcommand{\grouping}{\emph{grouping{}}}
\newcommand{\lfk}{\cite{lancichinetti-2009} {}}
\newcommand{\LFKNMI}{\ensuremath{\mbox{NMI}_{LFK}}}

\begin{abstract}
Given the increasing popularity of algorithms for overlapping clustering, in particular in social network analysis, quantitative measures are needed to measure the accuracy of a method.
Given a set of true clusters, and the set of clusters found by an algorithm, these sets of clusters must be compared to see how similar or different the sets are.
A normalized measure is desirable in many contexts, for example assigning a value of 0 where the two sets are totally dissimilar, and 1 where they are identical.

A measure based on normalized mutual information, \cite{lancichinetti-2009}, has recently become popular. We demonstrate unintuitive behaviour of this measure, and show how this can be corrected
by using a more conventional normalization. We compare the results to that of other measures, such as the Omega index \cite{collins1988omega}.

A C++ implementation is available online. \footnote{\url{https://github.com/aaronmcdaid/Overlapping-NMI}}
\end{abstract}

In a non-overlapping scenario, each node belongs to exactly one cluster. We are looking at overlapping, where a node could belong to many communities, or indeed to no clusters.
Such a set of clusters has been referred to as a \emph{cover} in the literature, and this is the terminology that we will use.

For a good introduction to our problem of comparing covers of overlapping clusters, see \cite{collins1988omega}.
They describe the Rand index, which is defined only for disjoint (non-overlapping) clusters, and then show how to extend it to overlapping clusters.
Each pair of nodes is considered and the number of clusters in common between the pair is counted. Even if a typical node is in many clusters,
it's likely that a randomly chosen pair of nodes will have zero clusters in common.
These counts are calculated for both covers and the Omega index is defined as the proportion of pairs for which the shared-cluster-count is identical,
subject to a correction for chance.

\section{Mutual information}

Meila \cite{MeilaNMI} defined a measure based on mutual information for comparing disjoint clusterings.
\citet{lancichinetti-2009} proposed a measure also based on mutual information, extended for covers.
This measure has become quite popular for comparing community finding algorithms in social network analysis.
It is this measure we are primarily concerned with there, and we will refer to it as \LFKNMI after the authors' initials.

We are proposing to use a different normalization to that used in \LFKNMI, but first we will define the non-normalized measure which is based very closely on that in \LFKNMI.
You may want to compare this to the final section of \citet{lancichinetti-2009}.

Given two covers, $X$ and $Y$, we must first see how to measure the similarity between a pair of clusters.
$X$ and $Y$ are matrices of cluster membership. There are $n$ objects.
The first cover has $K_X$ clusters, and hence $X$ is an
$n \times K_X$ matrix. $Y$ is an $n \times K_Y$ matrix.
$X_{im}$ tells us whether node $m$ is in cluster $i$ in cover $X$.

To compare cluster $i$ of the first cover to cluster $j$ of the second cover, we compare the vectors
$X_i$ and $Y_j$. These are vectors of ones and zeroes denoting which clusters the node is in.

\begin{itemize}
	\item $              { a = \sum_{m=1}^n [X_{im} = 0 \wedge  Y_{jm} = 0] } $
	\item $              { b = \sum_{m=1}^n [X_{im} = 0 \wedge  Y_{jm} = 1] } $
	\item $              { c = \sum_{m=1}^n [X_{im} = 1 \wedge  Y_{jm} = 0] } $
	\item $              { d = \sum_{m=1}^n [X_{im} = 1 \wedge  Y_{jm} = 1] } $
\end{itemize}

If $a+d = n$, and therefore $b=c=0$, then the two vectors are in complete agreement.

The lack of information between two vectors is defined:
\begin{align}
	H(X_i | Y_j) = & {} H(X_i , Y_j) - H(Y_j) \nonumber \\
	               = & {} h(a,n) + h(b,n) + h(c,n) + h(d,n) \nonumber \\
		         & {} - h(b+d,n) - h(a+c,n)
\end{align}
where $h(w,n) = -w \log_2 \frac{w}{n} $.

There is an interesting technicality here. Imagine a pair of clusters but where the memberships
have been defined randomly.
There is a possibility that there will be a small amount of mutual information, even in the situation where the two vectors are negatively correlated with each other.
In extremis, if the two vectors are near complements of each other, mutual information will be very high. We wish to override this and define that
there is zero mutual information in this case.
This is defined in equation (B.14) of \cite{lancichinetti-2009}.
We also use this restriction in our proposal.

\begin{equation}
	\begin{split}
	H^* & (X_i | Y_j) = \\
	& \left\{
		\begin{split}
			H(X_i | Y_j) \; & \mbox{~if} \; h(a,n) + h(d,n) \geq h(b,n) + h(c,n) \\
			h(c+d,n)+h(a+b,n) \;  & \mbox{~otherwise}
		\end{split}
	\right.
	\end{split}
	\label{eqnNoComplements}
\end{equation}

This allows us to compare vectors $X_i$ and $Y_j$, but we want to compare the entire matrices
$X$ and $Y$ to each other. We will follow the approximation used by \lfk here and
match each vector in $X$ to its best match in $Y$,

\begin{equation}
	H(X_i | Y) = \underset{j \in \{1,\dots K_Y \}}{\min} H^*(X_i | Y_j)
	\label{eqnBestMatch}
\end{equation}

then summing across all the vectors in $X$,

\begin{equation}
	H(X | Y) = \sum_{i \in \{1,\dots K_X \}} H(X_i | Y)
	\label{eqnSumVectors}
\end{equation}

$H(Y|X)$ is defined in a similar way to $H(X|Y)$, but with the roles reversed.
We will also need to define the (unconditional) entropy of a cover,
\begin{align*}
	H(X) & = \sum_{i=1}^{K_X} H(X_i) \\
	     & = \sum_{i=1}^{K_X} \left( h\left(\sum_{m=1}^n [ X_{im}=1 ] ,n\right) + h\left(\sum_{m=1}^n [ X_{im}=0 ] ,n\right)  \right) \;,
\end{align*}
where $\sum_{m=1}^n [ X_{im}=1 ]$ counts the number of nodes in cluster $i$,
end $\sum_{m=1}^n [ X_{im}=0 ]$ counts the number of nodes not in cluster $i$,

\section{Useful identities}

\begin{figure}
	\centering
\begin{tikzpicture}
	\draw (-1cm,0) circle(2cm);
	\draw (1cm,0) circle(2cm);
	\draw (0,0cm) node[fill=white] {$I(X:Y)$};
	\draw (1.8cm,0cm) node[fill=white] {$H(Y|X)$};
	\draw (-1.8cm,0cm) node[fill=white] {$H(X|Y)$};
	\path(1cm,0) ++  (60:2.2cm) node[rotate=-25] {$H(Y)$};
	\path(-1cm,0) ++ (120:2.2cm) node[rotate= 25] {$H(X)$};
\end{tikzpicture}
\caption{\label{figVenn} Mutual information and variation of information. The total information $H(X,Y) = H(X|Y) + I(X:Y) + H(Y|X)$. }
\end{figure}
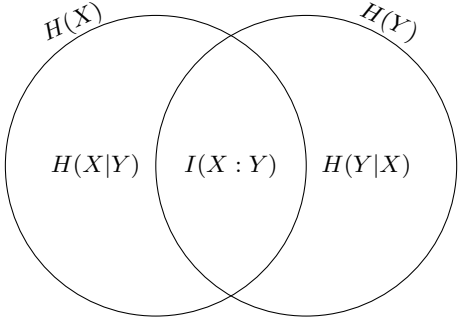

\cref{figVenn} gives us an easy way to remember the following useful identities, which
apply to any mutual information context.

\begin{align*}
	H(X) = & I(X:Y) + H(X|Y) \\
	H(Y) = & I(X:Y) + H(Y|X) \\
	H(X,Y) = & H(X) + H(Y|X) \\
	H(X,Y) = & H(Y) + H(X|Y) \\
	H(X,Y) = & \overbrace{I(X:Y)}^\text{mutual information} + \overbrace{H(X|Y) + H(Y|X)}^\text{variation of information}
\end{align*}

The first two equalities give us two definitions for the mutual information, $I(X:Y)$.
In theory, these should be identical, but due to the approximation used in \cref{eqnBestMatch}
they may be different. Therefore, we will use the average of the two.

\begin{equation}
	I(X:Y) := \frac12 \left[ H(X)-H(X|Y) + H(Y)-H(Y|X) \right]
	\label{eqnAverageOfTwo}
\end{equation}

We are now ready to discuss normalization, contrasting the method of \lfk
with our alternative.

\citet{lancichinetti-2009} define their own normalization of the \emph{variation of information},
\begin{equation}
	\frac12 \left(  \frac{ H(X|Y) }{ H(X) } + \frac{ H(Y|X) }{ H(Y) } \right)   \label{VOI}
\end{equation}
and hence their normalized mutual information is
\begin{equation}
	\LFKNMI
	= 1 - \frac12 \left(  \frac{ H(X|Y) }{ H(X) } + \frac{ H(Y|X) }{ H(Y) } \right) 
	\label{eqnLFKNMI}
\end{equation}

There are of course many ways to normalize a quantity such as the \emph{variation of information}.
Normalization typically involves division by a quantity $c$,
\begin{equation}
	\frac{ H(X|Y) + H(Y|X) } {c(X,Y)} \label{eqnNorms}
\end{equation}
where $c$ is a function of $X$ and $Y$ which is guaranteed to be greater than or equal to the numerator.
But \LFKNMI does not use a normalization of this standard form, instead using \cref{VOI}.

There is another aspect to the non-standard normalization used in \LFKNMI;
they insert an extra normalization factor into their definition of $H(X_i|Y_j)$.
But this is not the root cause of the problems we will describe, hence we will not dwell on it.
Our change is to remove all the normalization steps from their analysis and instead
use a more conventional normalization of the form of \cref{eqnNorms}.

\section{Unintuitive behaviour}
\label{secUnintuitive}
There are circumstances where \LFKNMI overestimates the similarity of two clusters.
We will show how an alternative normalization will fix these problems.

Imagine a cover $X$, and we are comparing it to a cover $Y$. Further, imagine $Y$ has only one
cluster ($K_Y=1$) and this cluster is identical to one of the clusters in $X$.
For large $K_X$, we would expect the normalized mutual information to be quite low.
An intuitive result would be approximately $\frac1{K_X}$.

However, $\LFKNMI(X,Y)$ will be at least $0.5$ in cases like this.
This is because $H(Y|X)$ will be zero bits
(the single cluster in $Y$ can be encoded with zero bits because it has a perfect match among the clusters of $X$)
and this will result in a contribution of $0.5$ to the \LFKNMI.

The other problematic example involves the power set. There are $n$ objects in total.
A cover involving every subset of the $n$ objects will create $2^n - 1$ clusters; we will ignore the empty subset.
This is the power set, which we denote as $p(n)$.

$\mbox{\LFKNMI}(X, p(n))$ will again be slightly greater than $0.5$.
This is because every cluster in $X$ will
have a perfect match in $p(n)$ and this will result in $H(X|p(n)) = 0$.

In both these examples \LFKNMI{} gives a score slightly above $0.5$. The intuitive behaviour
in these cases would be for a similarity score close to $0$.
We will demonstrate this behaviour in our experiments in \cref{secEval}

When we remove the
normalization from \LFKNMI, and instead use a more conventional normalization strategy
\cref{eqnNorms}, we will find more intuitive behaviour.

\section{normalization}
\label{secNormalization}

\begin{figure}[h!]
\includegraphics[width=0.5\textwidth]{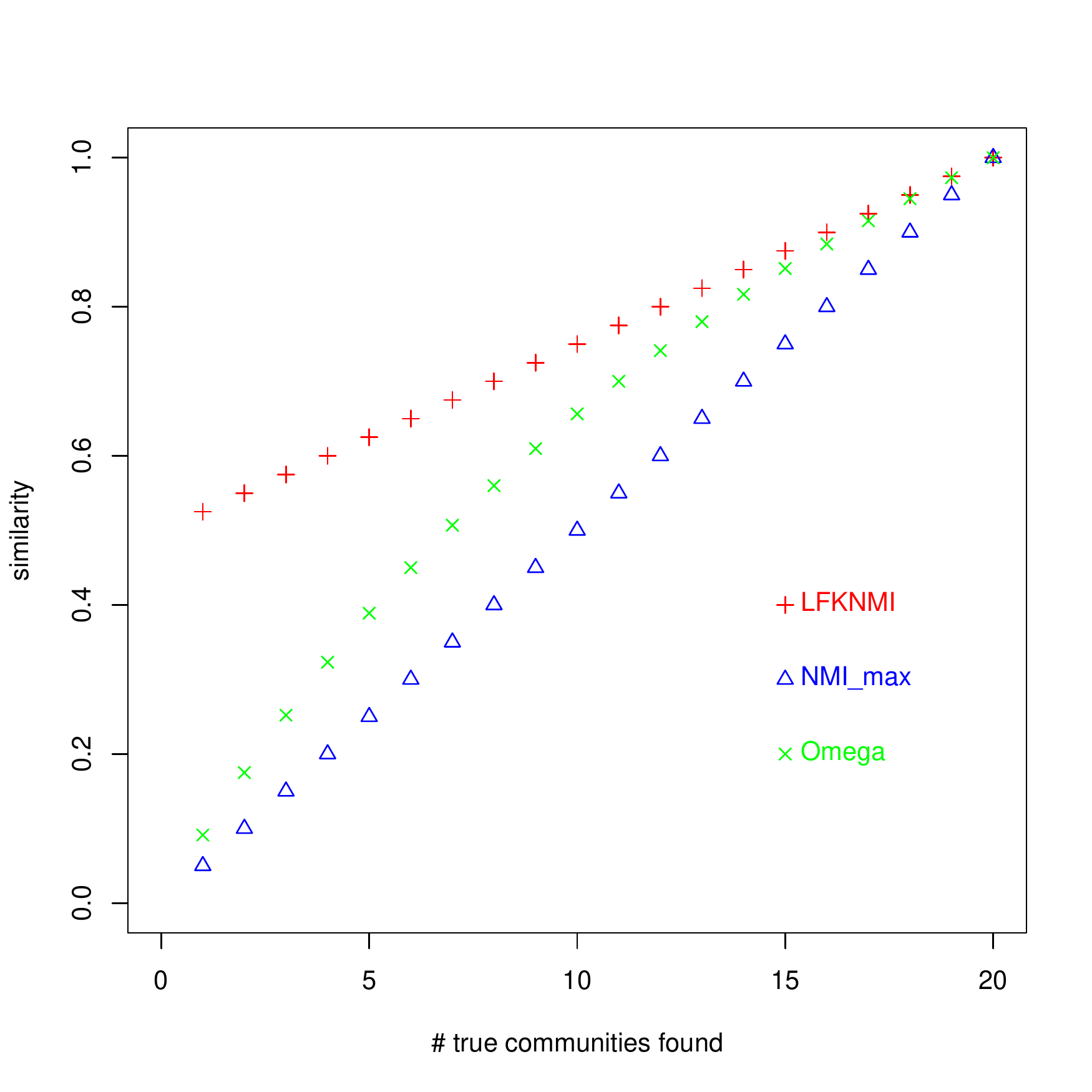}
\caption{As more communities are found, the scores of \LFKNMI and $\text{NMI}_{max}$ increase. For a small number of communities found, the intuitive result is a small value, and this is the behaviour of our proposed measure.}
\label{fig1to20}
\end{figure}

Typically a normalization will involve a simple division of the absolute quantity
by a quantity which is gauranteed to be an upper bound, giving us a number between
zero and one.

The following sequence of inequalities from \citet{VinhEppsBailey} provide possibilities for normalization.

\begin{equation}
	\begin{split}
	I(X:Y) \leq & \min(H(X),H(Y)) \\
	       \leq & \sqrt{H(X),H(Y)} \\
	       \leq & \frac12 \left(H(X)+H(Y)\right) \\
	       \leq & \max(H(X),H(Y)) \\
	       \leq & H(X,Y)
	\end{split}
\end{equation}

Any of the five expressions on the right can be used, and \cite{VinhEppsBailey} suggest
a measure based on $\max(H(X),H(Y))$. The Normalized Information Distance is recommended
\begin{equation*}
	d_{max} = 1 - \frac{ I(X,Y) }{ \max(H(X),H(Y)) }
\end{equation*}
where zero means perfect similarity and one means dissimilarity. We want a measure with
the opposite behaviour, so we'll use the corresponding normalized mutual information

\begin{equation}
	NMI_{max} = \frac{ I(X:Y) }{ \max(H(X),H(Y)) }
\end{equation}
where $I(X:Y)$ is as defined in \cref{eqnNoComplements,eqnBestMatch,eqnSumVectors,eqnAverageOfTwo}

This can also be understood with reference to \cref{figVenn}. The problem with \LFKNMI{} arises
when one cover is more complicated than the other, for example if one cover has many
more clusters than the other cover.
This corresponds to one circle in \cref{figVenn} being much larger than the other.
In both the unintuitive examples mentioned in \cref{secUnintuitive}, we will find that
one of the circles will be much larger than the other and that the overlap between
the two circles will be quite large, almost the full size of the smaller circle.
As a result, one of the terms inside the brackets in \cref{eqnLFKNMI}
will be small and will bring the \LFKNMI to 0.5.

\section{evaluation}
\label{secEval}

See \cref{fig1to20}. There are 200 nodes, divided into 20 communities. Each community has 10 nodes and they do not overlap.
We fix one of our covers, $X$, to be the full set of twenty communities. $Y$ contains a subset of these communities.
As we go from left to right, the number of communities in $Y$ increases from 1 to 20.

The communities in $Y$ are perfect copies of communities in $X$. Therefore, $X=Y$ when all 20 communities are used.
We see this in \cref{fig1to20} at the right, where both measures report an NMI of $1.0$.

This plot confirms the unintuitive behaviour of \LFKNMI when few communities are found.
On the left of the plot, when $Y$ has only one community, the score is $0.5$.

The linear relationship of our NMI$_{max}$, going from 0 to 1 as the number of communities in $Y$ increases, is intuitive.

\section{conclusion}

We have identified unintuitive behaviour in the version of NMI proposed by \lfk . We have identified the root
cause of the behaviour and shown how the use of a conventional normalization can lead to more intuitive behaviour.

A simple experiment was performed to confirm the existence of the unintuitive behaviour and demonstrate
the more intuitive behaviour.

There are a variety of normalized measures to measure the similarity of covers.
There is no unique set of evaluation criteria to decide on the best, but we suggest that our measure
is the most intuitive definition based on normalized mutual information.

\section{Acknowledgements}
This work is supported by Science Foundation Ireland under
grant 08/SRC/I1407: Clique: Graph and Network Analysis Cluster.


\bibliographystyle{unsrtnat}
\bibliography{community}
\end{document}